\documentclass[5p,preprint]{elsarticle}
\biboptions{sort&compress}
\usepackage{lineno,isotope,mathrsfs}
\usepackage{amsmath,bbold,amssymb,epsfig,bm,mathrsfs,feynmp}
\usepackage{color,slashed,nicefrac,exscale,multirow,soul}
\usepackage{times,txfonts,xcolor,graphicx,gensymb,tikz}
\usepackage[normalem]{ulem}
\usepackage{epstopdf}
\usepackage[colorlinks=true,linkcolor=blue,citecolor=blue]{hyperref}
\begin{document}
\begin{frontmatter}
\title{Rapidly rotating $\Delta$-resonance-admixed hypernuclear
compact stars}
\author[a,b,c]{Jia Jie Li}
\ead{jiajieli@itp.uni-frankfurt.de}
\author[d,e]{Armen Sedrakian}
\ead{sedrakian@fias.uni-frankfurt.de}
\author[f,g]{Fridolin Weber}
\ead{fweber@sdsu.edu, fweber@ucsd.edu}
\address[a]{School of Physical Science and Technology, Southwest University, Chongqing 400700, China}
\address[b]{Institute of Modern Physics, Chinese Academy of Sciences,
Lanzhou 730000, China}
\address[c]{Institute for Theoretical Physics, J. W. Goethe University,
D-60438 Frankfurt am Main, Germany}
\address[d]{Frankfurt Institute for Advanced Studies, D-60438
Frankfurt am Main, Germany}
\address[e]{Institute of Theoretical Physics, University of Wroclaw,
50-204 Wroclaw, Poland}
\address[f]{Department of Physics, San Diego State University,
5500 Campanile Drive, San Diego, California 92182, USA}
\address[g]{Center for Astrophysics and Space Sciences,
University of California at San Diego, La Jolla, California 92093, USA}
\begin{abstract}
We use a set of hadronic equations of state derived from covariant
density functional
theory to study the impact of their high-density behavior on the properties
of rapidly rotating $\Delta$-resonance-admixed hyperonic compact stars.
In particular, we explore systematically the effects of variations of the
bulk energy isoscalar skewness, $Q_{\mathrm{sat}}$, and the symmetry energy
slope, $L_{\mathrm{sym}}$, on the masses of rapidly rotating compact stars.
With models for equation of state satisfying all the modern astrophysical
constraints, excessively large gravitational masses of around
$2.5 \, M_{\odot}$ are only obtained under three conditions: (a) strongly
attractive $\Delta$-resonance potential in nuclear matter, (b) maximally
fast (Keplerian) rotation, and (c) parameter ranges
$Q_{\mathrm{sat}}\gtrsim500$ MeV and $L_{\mathrm{sym}}\lesssim50$
MeV. These values
of $Q_{\mathrm{sat}}$ and $L_{\mathrm{sym}}$ have a rather small
overlap with a large
sample (total of about 260) parametrizations of covariant nucleonic density
functionals. The extreme nature of requirements (a)-(c) reinforces the
theoretical expectation that the secondary object involved in the GW190814
event is likely to be a low-mass black hole rather than a supramassive
neutron star.
\end{abstract}
\end{frontmatter}
%
\section{Introduction}
\label{sec:intro}

In recent years there has been a surge of experimental information on the
integral parameters of neutron stars, mostly in the form of constraints
coming from their observations in gravitational and electromagnetic waves.
Among these is the first detection of gravitational waves from the binary
neutron star inspiral event GW170817 by the LIGO--Virgo Collaboration which
constrained the tidal deformability of a canonical 1.4\,$M_{\odot}$ mass
neutron star and thus the equation of state (EoS) of dense matter
at a few times nuclear saturation density~\cite{LIGO_Virgo2017a,LIGO_Virgo2017b,LIGO_Virgo2017c}.
These upper bounds suggest that the EoS of stellar matter at such (intermediate)
densities is medium-soft~\cite{LIGO_Virgo2018a,LIGO_Virgo2018b}.

A direct astrophysical lower bound of
$2.14^{+0.10}_{-0.09}\,M_{\odot}$ (68.3\% credibility interval) on the maximum mass of a neutron star was recently obtained
from the measurement of the millisecond pulsar PSR J0740+6620~\cite{Cromartie2019}.
The analysis of the GW170817 event was used to derive an
approximate upper limit on the maximum mass. By combining
gravitational waves and electromagnetic signals with numerical relativity
simulations, the maximum mass was found to be in the range of 2.15 to
$2.30\, M_{\odot}$~\cite{Margalit2017,Shibata2017,Ruiz2018}. The
quasi-universal
relations that describe neutron stars and models of kilonovae were used
to draw a similar bound on the maximum mass~\cite{Rezzolla2018}. Combining
the lower and upper bounds quoted above, it follows that the maximum mass
of a neutron star is in the $2.1$-2.3~$M_{\odot}$ range.

Furthermore, estimates of the mass and radius of the isolated 205.53 Hz
millisecond pulsar PSR J0030+0451 were reported from the analysis of the
NICER data of the thermal X-ray waveform from this object in 2019~\cite{Riley2019,Miller2019}.
The predicted radius and mass ranges of $R=13.02^{+1.24}_{-1.06}$ km and
$M=1.44^{+0.15}_{-0.14} \, M_{\odot}$ (68.3$\%$ credibility
interval)~\cite{Miller2019}
and the similar results by Ref.~\cite{Riley2019}, exclude both ultra-soft
as well as ultra-stiff behavior of the EoS at intermediate densities.
In particular, the relativistic (covariant) density functional based models, which predict somewhat larger radii appear to be consistent with the data if
the effects of heavy baryons such as hyperons and/or $\Delta$-resonances
are taken into account~\cite{Bonanno2012,Weissenborn2012a,Weissenborn2012b,Colucci2013,Dalen2014,Oertel2015,Chatterjee2016,Katayama2015,Fortin2016,Fortin2017,Chenyj2007,Drago2014,Caibj2015,Zhuzy2016,Sahoo2018,Kolomeitsev2017,Lijj2018b,Lijj2019a,Ribes2019,Lijj2020,Raduta2020}.

Very recently the LIGO--Virgo Collaboration observed gravitational waves
from a compact binary coalescence with an extremely asymmetric mass ratio
of involved compact object: the primary black hole mass is $22.2$-$24.3
\, M_{\odot}$ whereas the secondary mass is $2.50$-$2.67\, M_{\odot
}$~\cite{LIGO_Virgo2020}.
The mass of the latter object falls into the so-called ``mass-gap''
$2.5\, M_{\odot}\lesssim M\lesssim5\, M_{\odot}$ where no compact object
had ever been observed before. The absence of electromagnetic counterpart
and measurable tidal effects has left the nature of this compact object
open to interpretation. In particular, the interesting question arises
as to whether the light companion is the most massive neutron star or the
lightest black hole discovered to date. Several authors have addressed
this issue suggesting that we are dealing with an extremely rapidly rotating
nucleonic compact star
\cite{Most2020,Zhangnb2020,Tsokaros2020,Tews2020}. Rapid rotation is a
critical prerequisite of these scenarios, as it allows to increase a neutron
star's mass by around $\sim20\%$~\cite{Weber1992,Cook1994,Paschalidis2016}.
It was also found that static (i.e., non-rotating) nucleonic EoS models can
indeed generate massive stars with mass $M \gtrsim2.5\, M_{\odot}$, but
some of them are not compatible with constraints obtained from
GW170817~\cite{Fattoyev2020,Margueron2018b}.
A connection of the light companion in the GW190814 event with hyperonization
in dense matter was addressed by us in Ref.~\cite{Sedrakian2020} using
the well-calibrated DD-ME2 functional and its extension to the hypernuclear
sector. As pointed out in this paper, the compact star interpretation of
the light companion in GW190814 is in tension with hypernuclear stellar
models even in the case of maximal Keplerian rotation. In the present work,
we extend this study two-fold. First, we consider in detail the
$\Delta$-resonance admixture to the baryonic octet and study the sensitivity
of the results on the $\Delta$-potential in nuclear matter within the
set-up of our previous work~\cite{Lijj2018b}. Secondly, we study the
sensitivity
of the results with respect to variations of the (not well-constrained)
high-density behavior of the nucleonic density functional. To do so we use the well-known
Taylor expansions of the bulk and symmetry energies (see for
example~\cite{Margueron2018a,Zhangnb2018})
given by
%
%
\begin{eqnarray}
\label{eq:Taylor_expansion}
E(\chi, \delta) &\simeq& E_{\text{sat}} + \frac{1}{2!}K_{\text{sat}}
\chi^{2} + \frac{1}{3!}Q_{\text{sat}}\chi^{3}
\nonumber
\\
& & + E_{\text{sym}}\delta^{2} + L_{\text{sym}}\delta^{2}\chi+ {
\mathcal{O}}(\chi^{4},\chi^{2}\delta^{2}),
\end{eqnarray}
where $\chi=(\rho-\rho_{\mathrm{sat}})/3\rho_{\mathrm{sat}}$,
$\delta= (\rho_{n}-\rho_{p})/\rho$, $\rho_{n/p}$ are the neutron/proton
densities, and $\rho_{\mathrm{sat}}$ is the nuclear saturation density. The
first line
in the expansion \eqref{eq:Taylor_expansion} contains the characteristic
terms of the isoscalar channel, which are the saturation  energy
$E_{\mathrm{sat}}$, incompressibility $K_{\mathrm{sat}}$, and skewness
$Q_{\mathrm{sat}}$. The second line contains the characteristic
quantities of
the isovector channel, namely the symmetry energy
$E_{\mathrm{sym}}$ and its slope parameter $L_{\mathrm{sym}}$. Our
focus here will
be on the ``higher-order terms'' $Q_{\mathrm{sat}}$ and $L_{\mathrm{sym}}$ as these
are not well-determined so far.

Earlier, in Ref.~\cite{Zhangnb2020} the authors considered such an expansion
in the context of GW190814 being a fast-spinning neutron star, but without
an explicit reference to the particle content of the underlying model.
Indeed, expansions like~\eqref{eq:Taylor_expansion} can predict only the
amount of isospin in the matter, but are agnostic to its particle content
(unless one assumes that only neutrons and protons are present). Even less
informative on the particle content are the models which employ constant
speed-of-sound EoS~\cite{Alford2013,Zdunik2013,Alford2017} or piece-wise
polytropic EoS~\cite{Tews2020,Raaijmakers2018}, and such approaches cannot
be applied to study hypernuclear and/or $\Delta$-admixed matter. To gain
an access to the particle content of the star, we map the EoS given by
the expansion~\eqref{eq:Taylor_expansion} for each set of parameters
$Q_{\mathrm{sat}}$ and $L_{\mathrm{sym}}$ to a nucleonic density
functional, then
we take into account hyperons and $\Delta$-resonances with the parameters
tuned to the most plausible hyperon/resonance potentials extracted from
nuclear data.

In closing, we mention that an interesting physical possibility of the
behavior of superdense matter, which is not being studied here, concerns
the transition from hadronic to deconfined quark matter; for recent discussions
of this topic, see Refs.~\cite{Alford2017,Christian2020,Pereira2020,Ferreira2020,Blaschke2020,Bauswein2020,Lijj2020}.
This possibility in the present context of GW190814 event was discussed
in Refs.~\cite{Tan2020,Dexheimer2020}.

The paper is organized as follows. In Sec.~\ref{sec:theor_model} we briefly
review the key features of the covariant density functional (CDF) model
for hadronic matter. Particular attention is paid to the expansion coefficients
$Q_{\mathrm{sat}}$ and $L_{\mathrm{sym}}$ for nucleonic matter. This
is followed
in Sec.~\ref{sec:results} by a discussion of the bulk properties (in particular
maximal possible masses) of compact star models computed for a broad collection
of EoS identified in terms of $Q_{\mathrm{sat}}$ and $L_{\mathrm
{sym}}$. The key
findings of our study and their implications for the interpretation of
the GW190814 event are summarized in Sec.~\ref{sec:conclusions}.

\section{CDF model for hadronic matter}
\label{sec:theor_model}

At supranuclear density, hyperonization becomes a serious possibility since
hyperons are energetically favored in the cores of neutron stars~\cite{Glendenning1985,Glendenning1991}.
The presence of hyperons entails a considerable softening of the EoS which
lowers the (maximum) masses of neutron stars. In particular, such stars
have maximum masses that are smaller than those of neutron stars based
on purely nucleonic EoS~\cite{Vidana2001,Weissenborn2012a,Weissenborn2012b,Colucci2013,Dalen2014,Oertel2015,Tolos2016,Lijj2018a,Lijj2018b,Lijj2019a}.
At present, the existence of new degrees of freedom in the cores of neutron
stars can neither be confirmed nor ruled out based on astrophysical observations
alone. Indeed, one can readily generate hypernuclear EoS supporting a
$2\,M_{\odot}$ compact star~\cite{Weissenborn2012a,Weissenborn2012b,Colucci2013,Dalen2014,Oertel2015,Katayama2015,Tolos2016,Lijj2018a,Lijj2018b,Lijj2019a}.
In particular, CDF-based models are versatile enough to generate hypernuclear
EoS supporting a $2\,M_{\odot}$ compact star by fitting the parameters
of the interactions in the hyperonic sector to hypernuclear data~\cite{Dalen2014,Tolos2016,Fortin2017,Lijj2018a}.
These models, however, predict relatively large radii and tidal deformabilities
for neutron stars with canonical masses of around $1.4\, M_{\odot}$, which
is disfavored by the GW170817 data~\cite{Lijj2019a,Lijj2019b}. This issue
can be resolved if excited baryon states, in particular the
$\Delta$-resonance, are taken into account in the treatment of
$\beta$-equilibrated compact star matter~\cite{Drago2014,Lijj2018b,Ribes2019}.
As shown in Refs.~\cite{Lijj2019a,Lijj2019b,Ribes2019}, including the
$\Delta$-resonance in hypernuclear CDF calculations leads to neutron star
masses and radii that are no longer at variance with the values inferred
for those quantities from the observations of GW170817.

Here, we use the standard form of the CDF in which Dirac baryons are coupled
to mesons with density-dependent couplings~\cite{Vretenar2005,Mengjie2006}.
The theory is Lorentz invariant and, therefore, preserves causality when
applied to high-density matter. The baryons interact via the exchanges
of $\sigma,\,\omega$, and $\rho$ mesons, which comprise the minimal
set of mesons necessary for a quantitative description of nuclear phenomena.
In addition, we consider two hidden-strangeness mesons ($\sigma^{\ast
},\, \phi$) which describe interactions between hyperons.

The Lagrangian of the theory is given by the sum of the free baryonic and
mesonic Lagrangians, which can be found in Refs.~\cite{Lalazissis2005,Oertel2015,Lijj2018a},
and the interaction Lagrangian which reads
%
%
\begin{eqnarray}
\label{eq:interaction_Lagrangian}
\mathscr{L}_{\text{int}} &=&\sum_{B} \bar{\psi}_{B}\Big
(-g_{\sigma B}
\sigma-g_{\sigma^{\ast}B}\sigma^{\ast}-g_{\omega B}\gamma^{\mu
}\omega_{\mu}-g_{\phi B}\gamma^{\mu}\phi_{\mu
}\nonumber
\\
& &-g_{\rho B}\gamma^{\mu}\vec{\rho}_{\mu}\cdot\vec{\tau
}_{B}\Big)
\psi_{B} + \sum_{D} (\psi_{B} \rightarrow\psi^{\nu}_{D}),
\end{eqnarray}
where $\psi$ stands for the Dirac spinors and $\psi^{\nu}$ for the
Rarita-Schwinger
spinors~\cite{Pascalutsa2007}. Index $B$ labels the particles of the spin-1/2
baryonic octet, which comprises nucleons $N\in\{n,p\}$ and hyperons
$Y\in\{\Lambda,\Xi^{0,-},\Sigma^{+,0,-}\}$, while index $D$ refers
to the spin-3/2 resonance quartet of $\Delta$'s (i.e.,
$\Delta\in\{\Delta^{++,+,0,-}\}$). The mesons couple to the baryonic
octet and the $\Delta$'s with the strengths determined by the coupling
constants $g_{mB}$ and $g_{mD}$, which are functions of the baryonic density,
$g_{mB(D)}(\rho)=g_{mB(D)}(\rho_{\text{sat}})f_{m}(r)$, where
$r=\rho/\rho_{\text{sat}}$. There are in total four free parameters (three
in isoscalar sector and one in isovector sector) for functions
$f_{m}(r)$, which allow one to adjust the characteristic terms for nucleonic
matter $K_{\mathrm{sat}}$, $Q_{\mathrm{sat}}$, $L_{\mathrm{sym}}$ in
expansion~\eqref{eq:Taylor_expansion}
and $\rho_{\text{sat}}$, see Ref.~\cite{Lijj2019b} for detailed discussion
of the flexibility of functions $f_{m}(r)$. This study also suggests that
one can generate a set of nucleonic CDF models by varying only
$Q_{\mathrm{sat}}$ or $L_{\mathrm{sym}}$ while keeping the
lower-order parameters
fixed.

The Lagrangian~\eqref{eq:interaction_Lagrangian} is minimal, as it does
not contain (a) the isovector-scalar $\delta$ meson~\cite{RocaMaza2011}
and (b) the $\pi$ meson and the tensor couplings of vector mesons to baryons
(both of which arise in the Hartree-Fock theory~\cite{Lijj2018a}). As shown
below in Sec.~\ref{sec:results}, a wide range of the mass-radius relations
can be generated by this Lagrangian which covers parameter space
comparable with the recent meta-modeling for realistic nucleonic
EoS~\cite{Margueron2018b}.
We note also that other spin-3/2 resonances (like $\Sigma^{-*}$ which
has a slightly heavier mass than $\Delta$) may also appear in dense matter.
However, their potentials in nuclear matter are unknown. We thus consider only
the lightest (non-strange) members of the baryon
$J^{3/2}$-decouplet.

%
\begin{figure}[tb]
\centering
\includegraphics[width = 0.80\hsize]{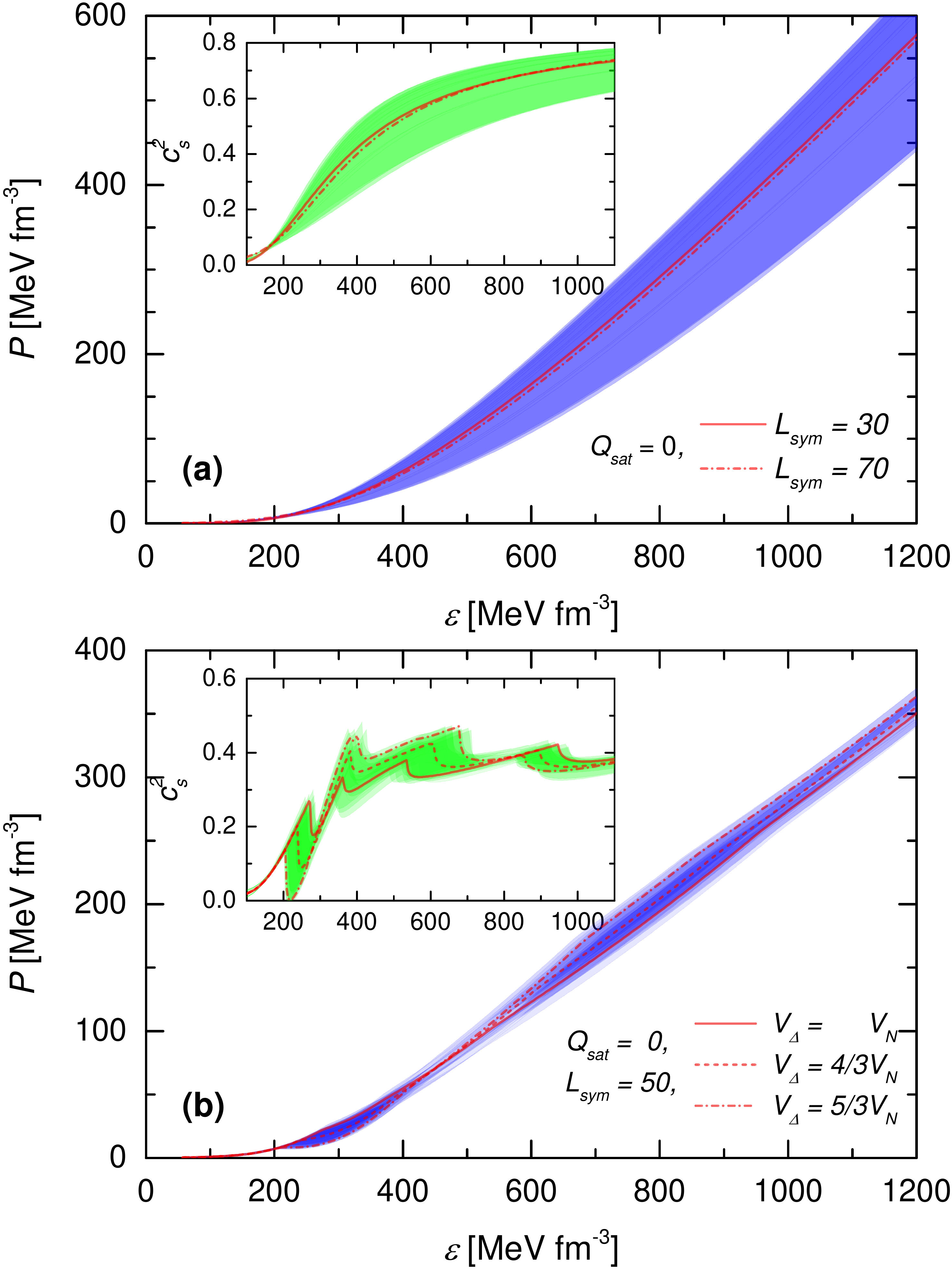}
\caption{EoS and the corresponding speed-of-sound squared for (a) purely
nucleonic and (b) $\Delta$-hyperon admixed stellar matter. In (a) the nucleonic EoS models are generated by varying the parameters $Q_{\mathrm{sat}}$
$\in[-600, 900]$~MeV and $L_{\mathrm{sym}}$ $\in[30, 70]$~MeV. The EoS with
$Q_{\mathrm{sat}}= 0$, $L_{\mathrm{sym}}= 30$ and 70~MeV are shown by solid and
dash-dotted lines for illustration. In (b) $\Delta$-admixed hyperonic matter EoS
are generated by varying the parameters $Q_{\mathrm{sat}}\in[300, 900]$~MeV,
$L_{\mathrm{sym}}\in[30, 70]$~MeV for values of $\Delta$-potential $V_{D}$ in
isospin symmetric nuclear matter $V_{D}/V_{N} = 1, 4/3$ and $5/3$, where $V_{N}$
is the nucleonic potential. The EoS models with $Q_{\mathrm{sat}}= 600,
L_{\mathrm{sym}}= 50$~MeV and three indicated values of $V_{D}$ are shown for
illustration.}
\label{fig:EOS}
\end{figure}

For our analysis below we adopt, as a reference, the DD-ME2
parametrization~\cite{Lalazissis2005}
which was calibrated to the properties of finite nuclei. This parametrization
has been tested on the entire nuclear chart with great success and agrees
with experimentally known bounds on the empirical parameters of nuclear
matter. In the hypernuclear sector, the vector meson-hyperon couplings
are given by the SU(6) spin-flavor-symmetric quark model, whereas the scalar
meson-hyperon couplings are determined by fitting them to the potentials
extracted from hypernuclear systems. For the resonance sector, the vector
meson-$\Delta$ couplings are chosen close to the meson-$N$ ones, whereas
the scalar meson-$\Delta$ couplings are determined by fitting them to
certain preselected potentials extracted from heavy-ion collisions and
the scattering of electrons and pions off nuclei (for an overview see
Refs.~\cite{Drago2014,Kolomeitsev2017,Lijj2018b,Raduta2020}).
Note that in this manner \textit{we assume that the hyperon and
$\Delta$ potentials scale with density the same way as the nucleonic
potentials,
and therefore their high-density behavior is inferred from that of the
nucleons}. This assumption has its justification in the quark substructure
of the constituents. However, first-principle computations that may support
our assumption is still lacking. See Refs.~\cite{Lijj2019a,Lijj2019b}
for details of the model.

The nuclear matter EoS can be characterized in terms of the double expansion,
shown in Eq.~\eqref{eq:Taylor_expansion}, around the saturation density
and the isospin symmetrical limit. In Refs.~\cite{Margueron2018a,Lijj2019b}
it has been shown that the gross properties of compact stars are very sensitive
to the higher-order empirical parameters of nuclear matter around the saturation
density, specifically to the isoscalar skewness $Q_{\mathrm{sat}}$ and
isovector
slope $L_{\mathrm{sym}}$. Note also that the low-order empirical parameters
are well constrained by physics of finite nuclei. The combined analysis
of terrestrial experiments and astrophysical observations predict a value
for the slope of symmetry energy $L_{\mathrm{sym}}= 58.7\pm28.1$~MeV \cite{Oertel2017}. The skewness $Q_{\mathrm{sat}}$ is highly
model dependent.
For example, non-relativistic Skyrme or Gogny models predict (predominantly)
negative $Q_{\mathrm{sat}}$ value~\cite{Dutra2012,Margueron2018a},
whereas relativistic
models predict both positive and negative $Q_{\mathrm{sat}}$
values~\cite{Dutra2014,Margueron2018a,Tongh2018}.
With this in mind, we vary $Q_{\mathrm{sat}}$ and/or $L_{\mathrm{sym}}$
individually
within a wide range and study their impact on the properties of compact stars,
by modifying (only!) the density-dependence of the functional at high density;
its well-tuned features at and around the saturation density remain fixed
as the defaults of DD-ME2~\cite{Lalazissis2005}, namely,
$K_{\mathrm{sat}}= 251.2$~MeV and $E_{\mathrm{sym}}= 32.3$~MeV.

Fig.~\ref{fig:EOS} shows the EoS and the corresponding speed-of-sound
squared for purely nucleonic and $\Delta$-admixed hyperonic stellar matter
for a range of the parameters $Q_{\mathrm{sat}}, L_{\mathrm{sym}}$
and values of
$\Delta$-potential $V_{D}$ in symmetric nuclear matter ($V_{D}/V_{N} =
1, 4/3$ and $5/3$, where $V_{N}$ is the nucleonic potential). All those
EoS models fulfill the constraints of $2\,M_{\odot}$
observations~\cite{Antoniadis2013,Cromartie2019}.
The results for $\Delta$-admixed hyperonic stellar matter with
$V_{D}/V_{N} < 1$ are not shown, since in this case,
the $\Delta$ population is rather small~\cite{Lijj2018b,Ribes2019}. In
Fig.~\ref{fig:EOS}\,(b) we illustrate also the EoSs of $\Delta$-admixed
hyperonic matter for three values of $\Delta$-potential. It is seen that
$\Delta$'s \textit{soften} the EoS at low densities which directly implies
smaller radii for not very massive members of the sequences. This effect
increases with the depth of the $\Delta$-potential, i.e., the larger is
the attractive the $\Delta$-potential, the smaller is the radius of the
intermediate-mass compact star (see Fig.~\ref{fig:MRL_Del} below).

\section{Results and discussions}
\label{sec:results}

\subsection{Nucleonic EoS models}
\label{sec3.1}

%
\begin{figure}[tb]
\centering
\includegraphics[width = 0.80\hsize]{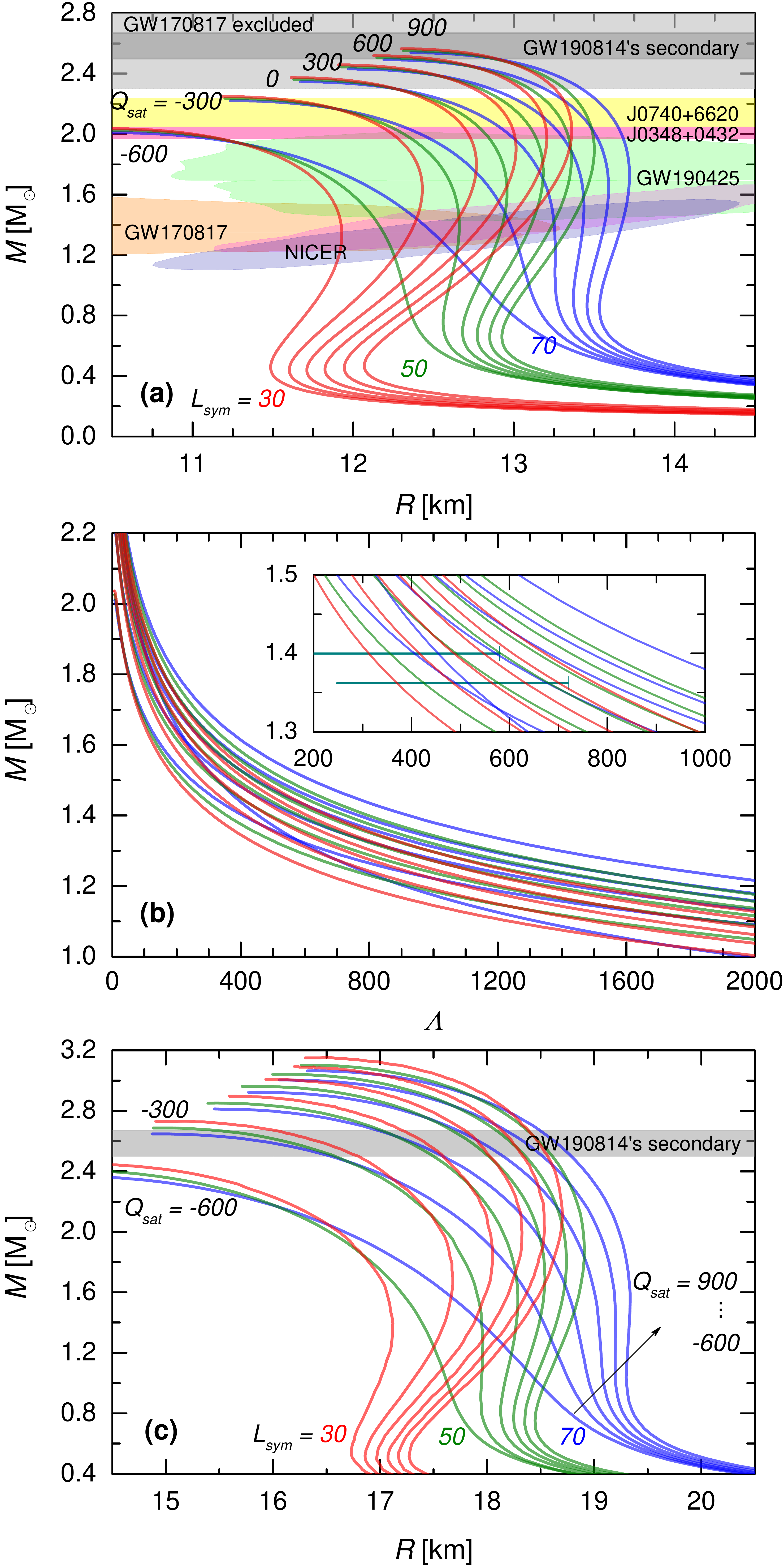}
\caption{Mass-radius (a) and mass-tidal deformability (b) relations of static
(i.e., non-rotating), purely nucleonic stellar configurations generated by
tuning the isoscalar skewness coefficient $Q_{\mathrm{sat}}$ and the slope of
symmetry energy $L_{\mathrm{sym}}$. Modern constraints from multi-messenger
astronomy are shown by the color regions (see text for details). (c) Same as
(a), but for rapidly rotating (Keplerian) sequences.}
\label{fig:MRL_Nuc}
\end{figure}

We first consider static (non-rotating) as well as rapidly rotating compact
stars made of purely nucleonic matter. Figs.~\ref{fig:MRL_Nuc}\,(a) and
(b) show the mass-radius and mass-tidal deformability relationships computed
for $Q_{\mathrm{sat}}$ values $-600$, $-300$, 0, 300, 600, and 900 MeV
(in that
order from left to right) and $L_{\mathrm{sym}}$ values of 30 (red curves),
50 (green curves), and 70 MeV (blue curves). Observational constraints
from multi-messenger astronomy are highlighted. These concern the masses
of PSR J0348+0432~\cite{Antoniadis2013} and PSR J0740+6620~\cite{Cromartie2019},
the compactness and tidal deformability constraints extracted from the
binary compact star mergers GW170817~\cite{LIGO_Virgo2018b,Coughlin2018,Kiuchi2019}
and GW190425~\cite{GW190425}, the mass and radius measurements for PSR
J0030+0451 by NICER~\cite{Riley2019,Miller2019}, and the mass of the secondary
component of GW190814~\cite{LIGO_Virgo2020}.

One sees from Fig.~\ref{fig:MRL_Nuc}\,(a) that compact stars with masses
of around $M\sim2.5\, M_{\odot}$ require nucleonic EoS models with large
and positive $Q_{\mathrm{sat}}$ values in the range
$Q_{\mathrm{sat}}\gtrsim600 $~MeV, where $L_{\mathrm{sym}}$ can be
30, 50, or 70
MeV. These EoS models, however, lead to
$12.9 \lesssim R_{1.4} \lesssim13.7$ km for the radius of a
$1.4\, M_{\odot}$ star, as can be read off from {Fig.~\ref{fig:MRL_Nuc}}\,(a),
and to tidal deformabilities $\Lambda_{1.4} \gtrsim700$ (Fig.~\ref{fig:MRL_Nuc}\,(b)),
both of which being at variance with GW170817 observation~\cite{LIGO_Virgo2018b}.
In fact the revised upper limit on $\Lambda_{1.4}$ is
$190^{+390}_{-120}$ (90\% credibility interval)~\cite{LIGO_Virgo2018b},
which does not overlap with $\Lambda_{1.4} \gtrsim700$. Furthermore,
we checked that the EoSs for symmetric nuclear matter computed with these
models are much stiffer than the range of admissible EoS deduced from studies
of heavy-ion collisions~\cite{Danielewicz2002}. A similar conclusion was
reached also in a recent work where the nonlinear CDF models were
used~\cite{Fattoyev2020}.

The only EoS models that lead to $\Lambda_{1.4}$ values compatible with
$\Lambda_{1.4} = 190^{+390}_{-120}$ are those computed for
($L_{\mathrm{sym}}=30$
MeV, $Q_{\mathrm{sat}}\lesssim300$ MeV), ($L_{\mathrm{sym}}=50$ MeV,
$Q_{\mathrm{sat}}\lesssim0$ MeV) and ($L_{\mathrm{sym}}=70$, MeV,
$Q_{\mathrm{sat}}\lesssim-300$ MeV), as can be deduced from the
curves shown
in the inset in Fig.~\ref{fig:MRL_Nuc}\,(b). All these combinations correspond
to $\Lambda_{1.4} \lesssim580$ MeV, the upper bound of inferred
$\Lambda_{1.4} = 190^{+390}_{-120}$. In summary, we conclude that the
low tidal deformability of a $1.4 \, M_{\odot}$ compact star inferred from
GW170817 makes it highly unlikely that the maximum mass of a static, nucleonic
neutron star could be as high as $\sim2.5\, M_{\odot}$.

Next, we turn to the maximally rotating stellar models shown in
Fig.~\ref{fig:MRL_Nuc}\,(c).
There exist several codes for computing configurations of rapidly rotating
compact stars, all of which are based on the iterative method of solution
of Einstein's equations~\cite{Nozawa1998,Cook1994} in axial symmetry for
any tabulated EoS. The method starts with a ``guess'' density
profile, integrates the stellar structure equations, thus obtaining a new
input density profile for the following iteration. This procedure is repeated
until convergence is achieved at each point of the spatial grid. In our
computations, we use the public domain RNS code\footnote{{www.gravity.phys.uwm.edu/rns/}.}
which implements this scheme. Each star shown in this figure rotates at
its respective (general relativistic) Kepler frequency, at which mass shedding
from the equator terminates stable rotation. There are other rotational
instabilities (like the $r$-modes) which set a tighter limit on stable
rotation than the Kepler frequency does. However, the Kepler frequency
is particularly interesting as it sets an absolute limit on rapid rotation,
and it also enables stars to carry the maximum amount of mass. From model
calculation it is known that the gravitational mass increase can be as
large as around $20\%$~\cite{Weber1992,Cook1994} compared to non-rotating
stars. As shown in Fig.~\ref{fig:MRL_Nuc}\,(c), almost all EoS models are capable of producing a compact star whose mass falls in the mass range
estimated for the secondary in GW190814. The only models that fail are
those based on very negative $Q_{\mathrm{sat}}$ values (e.g.,
$Q_{\mathrm{sat}}= -600$ MeV), independent of the value chosen for
$L_{\mathrm{sym}}$. The $Q_{\mathrm{sat}}= -300$ MeV
EoS models, which fail
to support a non-rotating $\sim2.5 \, M_{\odot}$ compact star
(Fig.~\ref{fig:MRL_Nuc}\,(a)),
now support stars with masses exceeding $2.6\,M_{\odot}$.

%
\begin{figure}[tb]
\centering
\includegraphics[width = 0.80\hsize]{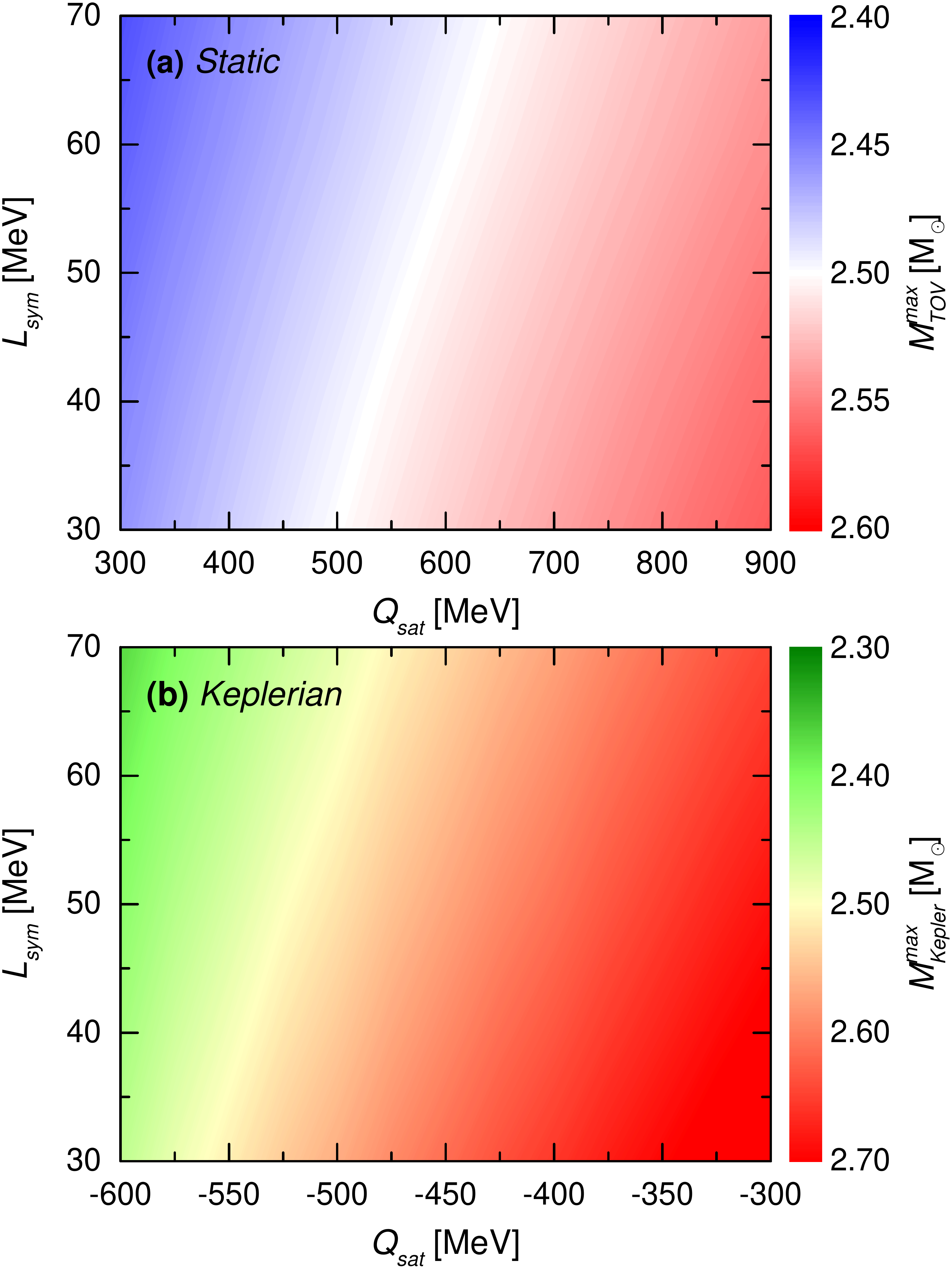}
\caption{The maximum masses of (a) static and (b) Keplerian purely nucleonic
stellar sequences (color coded column on the right) for a range of values
spanned by $Q_{\mathrm{sat}}$ and $L_{\mathrm{sym}}$. The
large-$Q_{\mathrm{sat}}$ and small-$L_{\mathrm{sym}}$ range corresponds to
compact stars with masses exceeding $2.5\,M_{\odot}$.}
\label{fig:MQL_Nuc}
\end{figure}

The dependence of the maximum masses for static and Keplerian models on
the relevant range of $Q_{\mathrm{sat}}$ and $L_{\mathrm{sym}}$
parameters for purely
nucleonic EoS models is shown in Fig.~\ref{fig:MQL_Nuc}. It allows one
to easily read-off the maximum masses predicted by any density functional
once its values for $Q_{\mathrm{sat}}$ and $L_{\mathrm{sym}}$ are known.

To summarize, (i) static nucleonic compact stars with masses up to
$2.5\, M_{\odot}$ can be obtained for $Q_{\mathrm{sat}}\gtrsim
600$~MeV, however,
the tidal deformability $\Lambda_{1.4}$ for such models is in tension
with the inference from GW170817; (ii) this tension is lifted if one assumes
that the unknown secondary in GW190814 is a rapidly rotating neutron star
composed of nucleonic matter~\cite{Most2020,Zhangnb2020,Tsokaros2020};
(iii) nevertheless, if $Q_{\mathrm{sat}}\leq-500$~MeV, the static
(Tolman-Oppen\-heimer-Volkoff)
maximum masses of the models are
$M^{\mathrm{max}}_{\mathrm{TOV}} \leq2.1 \, M_{\odot}$ and such
models are in agreement
with tidal deformability values $\Lambda_{1.4}$ derived from the GW170817
event and the radius values $R_{1.4}$ obtained from NICER's X-ray
observations~\cite{Riley2019,Miller2019}.
In this case the secondary of GW190814 must be a black hole.

\subsection{Hyperon-$\Delta$ admixed EoS models}
\label{sec:delta_admixed}

%
\begin{figure}[tb]
\centering
\includegraphics[width = 0.80\hsize]{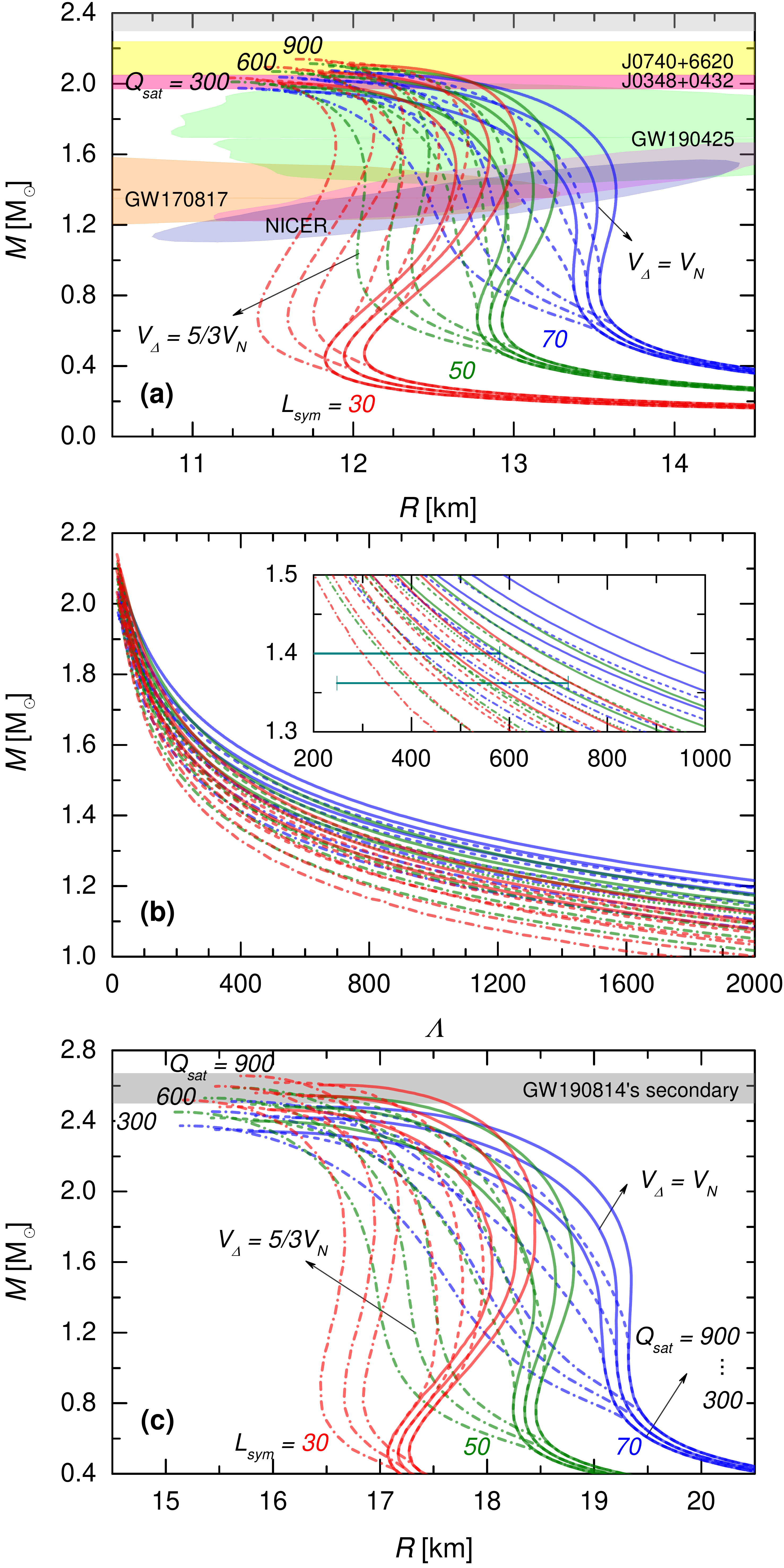}
\caption{Mass-radius (a) and mass-tidal deformability (b) relations of static
stellar configurations containing hyperon-$\Delta$-admixed matter, generated by
tuning the isoscalar skewness coefficient $Q_{\mathrm{sat}}$ and the slope of
symmetry energy $L_{\mathrm{sym}}$, and the $\Delta$-potential at nuclear
saturation density $V_{\Delta}/V_{N}=1$ (solid lines), 4/3 (dashed), and $5/3$
(dash-dotted). (c) Same as (a), but for rapidly rotating (Keplerian) sequences.}
\label{fig:MRL_Del}
\end{figure}
%
%
\begin{figure}[tb]
\centering
\includegraphics[width = 0.80\hsize]{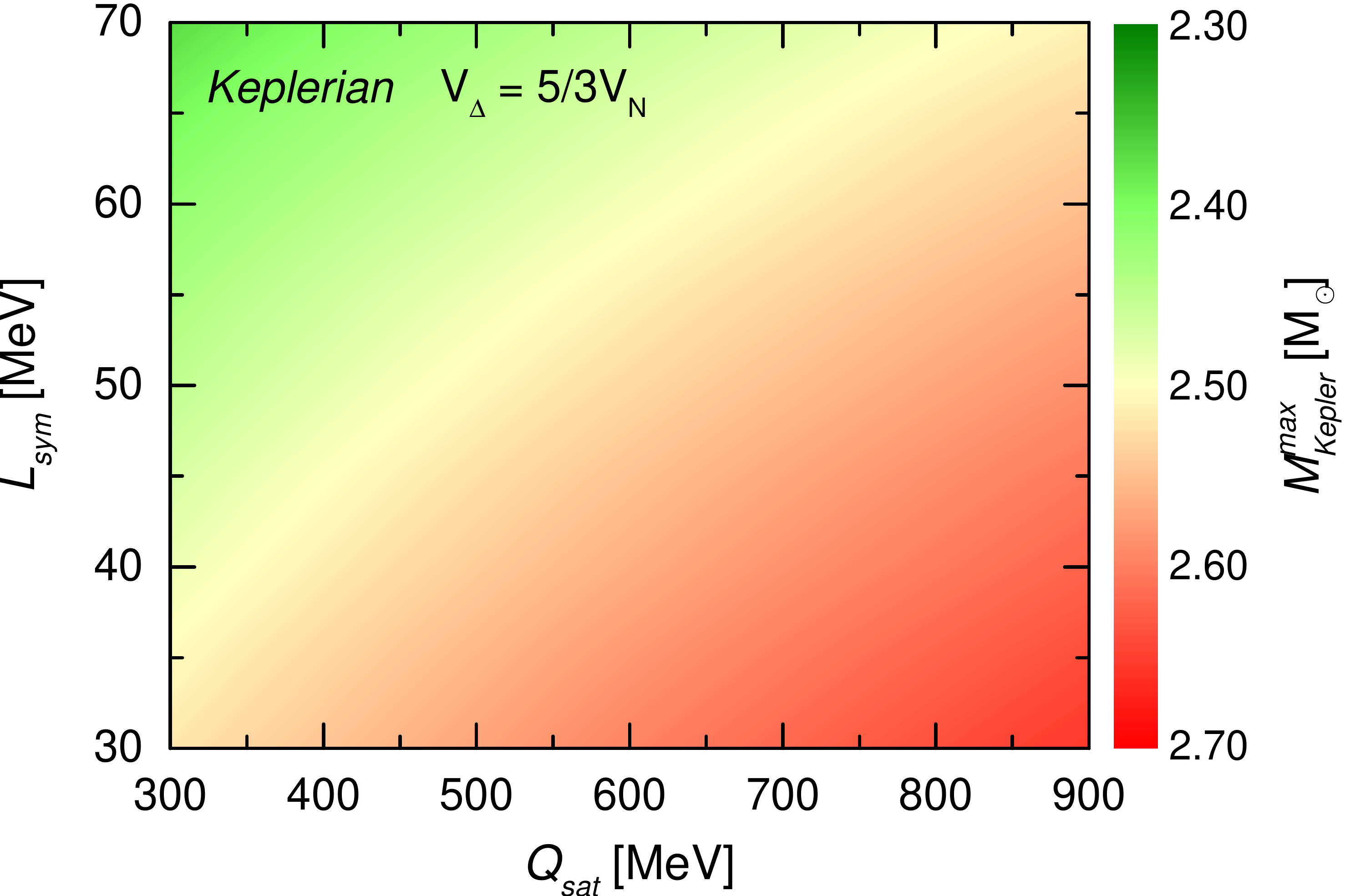}
\caption{The maximum masses of Keplerian sequences as a function parameter space
spanned by $Q_{\mathrm{sat}}$ and $L_{\mathrm{sym}}$. The $\Delta$-resonance
potential is fixed at the largest value $V_{\Delta}=5/3V_{N}$ considered in this
work. The large-$Q_{\mathrm{sat}}$ and small-$L_{\mathrm{sym}}$ range
corresponds to compact stars with masses exceeding $2.5\,M_{\odot}$.}
\label{fig:MQL_Del}
\end{figure}

Since the matter in the cores of compact stars is compressed to densities
several times higher than the density of atomic nuclei, the core composition
may contain substantial populations of hyperons and, as emphasized in several
recent papers, by $\Delta$'s too~\cite{Drago2014,Caibj2015,Zhuzy2016,Sahoo2018,Kolomeitsev2017,Lijj2018b,Lijj2019a,Ribes2019,Spinella2020,Malfatti2020}.
The possible presence of $\Delta$'s in the cores of neutron stars has
not been considered for years since the early CDF calculations did show
that the $\Delta$-resonance would appear at densities too high to be reached
in the cores of compact stars~\cite{Glendenning1985}. However, calculations
based on more sophisticated microscopic models and/or tighter constraints
on the model parameters
~\cite{Drago2014,Caibj2015,Zhuzy2016,Sahoo2018,Kolomeitsev2017,Lijj2018b,Lijj2019a,Ribes2019,Spinella2020,Malfatti2020,Schurhoff2010}
show that $\Delta$'s could make up a large fraction of the baryon population
in neutron star matter and could also have a significant effect on the
radii of compact stars~\cite{Schurhoff2010,Lijj2018b,Ribes2019}.

In Figs.~\ref{fig:MRL_Del}\,(a) and (b) the mass-radius and mass-tidal
deformability relations computed for $\Delta$-admixed hyperonic EoS models
are shown for $Q_{\mathrm{sat}}$ values ranging from 300 to 900 MeV,
$L_{\mathrm{sym}}$ values of 30, 50 and 70 MeV, and different values
for the
$\Delta$-potential $V_{\Delta}$ at nuclear saturation density. As can be
seen, to support a compact star with a gravitational mass of about
$2\, M_{\odot}$, containing hyperons and $\Delta$'s in its core,
$Q_{\mathrm{sat}}$ needs to be at least as large as $\sim300$~MeV.
The maximum
possible mass of the static stellar sequence is
$M^{\mathrm{max}}_{\mathrm{TOV}} \simeq2.2\, M_{\odot}$.

Imposing the $\Lambda_{1.4} = 190^{+390}_{-120}$ constraint on the EoS,
it follows from Fig.~\ref{fig:MRL_Del}\,(b) that all hyperon-$\Delta$-admixed EoS models are consistent with this constraint
if the $\Delta$-potential is assumed to be $V_{\Delta}/V_{N}=5/3$, independent
of the particular choices of $Q_{\mathrm{sat}}$ and $L_{\mathrm
{sym}}$. The situation
is strikingly different for $V_{\Delta}/V_{N}=1$ in which case only
$Q_{\mathrm{sat}}=300$ and 600 MeV are allowed for $L_{\mathrm
{sym}}=30$ MeV. For
$L_{\mathrm{sym}}=70$ MeV none of the three $Q_{\mathrm{sat}}$
values leads
to tidal deformabilities that are in agreement with
$\Lambda_{1.4} \leq580$.

{Fig.~\ref{fig:MRL_Del}}\,(c) shows the mass-radius relationships of maximally
rotating (Keplerian) stellar models computed for our collection of
$\Delta$-admixed hyperonic EoS models. As can be seen, the rotation at
the mass shedding limit increases the maximum-possible gravitational mass
to values in the range of
$2.4\, M_{\odot} \lesssim M^{\mathrm{max}}_{\mathrm{Kepler}}
\lesssim2.7\, M_{
\odot}$, depending on the $Q_{\mathrm{sat}}$ and $L_{\mathrm{sym}}$ values and the
depth of the $\Delta$-potential. The largest values for
$M^{\mathrm{max}}_{\mathrm{Kepler}}$ are obtained for $Q_{\mathrm
{sat}}\geq600$~MeV,
$L_{\mathrm{sym}}\leq50$~MeV, and $V_{\Delta}/V_{N} = 5/3$. All
these models
for the EoS lead to masses that are consistent with the mass estimated
for the stellar secondary of the GW170817 event.

The dependence of maximum masses of the Keplerian models on the range of
$Q_{\mathrm{sat}}$ and $L_{\mathrm{sym}}$ parameters in the case
$V_{\Delta} =5/3 V_{{N}}$ is shown in Fig.~\ref{fig:MQL_Del}. We note that
the range of $Q_{\mathrm{sat}}$ value extracted from our analysis has
a rather
small overlap with the ones extracted from large samples of non-relativistic
and relativistic density functionals~\cite{Dutra2012,Dutra2014}. We thus
conclude that for the secondary object in GW190814 to be a compact star
featuring heavy baryons requires several extreme assumptions, which apart
from maximally rapid rotation, requires large values for the
$\Delta$-resonance potential in nuclear matter and combinations of
$Q_{\mathrm{sat}}$ and $L_{\mathrm{sym}}$ that fall outside the range
covered by
all known density functionals, except DD-ME2~\cite{Lalazissis2005} and
a few newly proposed functionals~\cite{Taninah2020,Fattoyev2020}. These
findings support the theoretical expectation that the secondary stellar
object involved in the GW190814 event is a low-mass black hole rather than
a supermassive neutron star.

Finally, it should be mentioned that in this work the vector meson-hyperon
couplings are given by the SU(6) spin-flavor symmetric quark model. If
one fixes the couplings according to the more general SU(3) flavor symmetry,
the maximum mass of static compact stars would increase by about 10\%
\cite{Weissenborn2012b,Lijj2018a}.
However, we anticipate that modification of vector meson-hyperon couplings
will not change our main conclusion about the nature of GW190814's secondary
member.

\section{Summary and conclusions}
\label{sec:conclusions}

In this work, we have investigated properties of non-rotating as well as
rapidly rotating compact stars with and without $\Delta$-resonance-admixed
hyperonic core compositions. The corresponding models for the EoS are generated
with  covariant density functional theory. The high-density behavior
of nucleonic EoS is quantified in terms of the isoscalar skewness coefficient
$Q_{\mathrm{sat}}$ and the isovector slope coefficient $L_{\mathrm
{sym}}$. The hyperon
potentials are tuned to the most plausible potentials extracted from
hypernuclear
data. The $\Delta$-potential in nuclear matter is taken to be in
the range $1 \leq V_{\Delta}/V_{N} \leq5/3$, as no consensus has been reached
yet on its magnitude. The density-dependences of the hyperon- and
$\Delta$-meson couplings are assumed to be the same as those of nucleons.

We found that purely nucleonic models for the EoS can accommodate compact
stars as massive as $M \simeq2.5\, M_{\odot}$, but only if the isoscalar
skewness coefficient $Q_{\mathrm{sat}}\gtrsim600$~MeV. These EoS
models, however,
lead to tidal deformabilities for a $1.4 \, M_{\odot}$ star that conflict
with observation and are thus ruled out as valid EoS models. To resolve
the tidal deformability issue, one must have
$Q_{\mathrm{sat}}\lesssim300$~MeV. The problem that arises from these
EoS models,
however, is that they then no longer support a $2.5\, M_{\odot}$ star and
thus qualify either. The maximal possible rotation rate at the mass shedding
limit resolves this issue as it pushes the masses of most (with the exception
of $Q_{\mathrm{sat}}\lesssim-550$~MeV) stellar sequences up to the
$\sim2.5\, M_{\odot}$ mass range. This confirms the earlier findings~\cite{Most2020,Zhangnb2020,Tsokaros2020}
that a rapidly, uniformly rotating compact star made of purely nucleonic
matter could have been the secondary stellar object involved in the GW190814
event.

Taking hyperon and $\Delta$-resonance populations into account our EoS
models reduces the masses of compact stars. In particular, the
maximal masses of non-rotating stars are reduced to
$2.0 \, M_{\odot}\lesssim M \lesssim2.2 \, M_{\odot}$ if
$Q_{\mathrm{sat}}\gtrsim300$~MeV. So none of these models comes even close to the
$2.5\, M_{\odot}$ constraint set by GW190814. This is different if rapid
rotation at the mass shedding frequency is considered. In this case, the
stellar models computed for a strongly attractive $\Delta$-potential in
nuclear matter of $V_{\Delta}/V_{N} =5/3$ reach the $2.5\, M_{\odot}$ mass
limit rather comfortably. The situation is strongly depending on the
$Q_{\mathrm{sat}}$ and $L_{\mathrm{sym}}$ values, as graphically
illustrated in {Fig.~\ref{fig:MQL_Del}}.
The smallest value for $Q_{\mathrm{sat}}$ is $ Q_{\mathrm
{sat}}\approx300$ MeV for
$L_{\mathrm{sym}}= 30$~MeV, while $Q_{\mathrm{sat}}\approx900$~MeV for
$L_{\mathrm{sym}}= 70$~Mev. We note that all the valid EoS models in
this figure
lead to $R_{1.4}$ and $\Lambda_{1.4}$ values that are in agreement with
observation. For EoS models computed with $\Delta$-potential
$V_{\Delta}/V_{N} =1$, the agreement is either only marginal or can not
be reached at all. The combinations required for $Q_{\mathrm{sat}}$ and
$L_{\mathrm{sym}}$ lie outside the range covered by presently known
non-relativistic
and relativistic nuclear density functionals. A few exceptions to this
are the functionals with values $Q_{\mathrm{sat}}\gtrsim 500$ MeV and
the possibility
of $M^{\mathrm{max}}_{\mathrm{Kepler}}/M_{\odot} \gtrsim2.5$.

To summarize, current valid EoS models which account for $\Delta$-admixed
hyperonic matter in the cores of compact stars imply that the secondary
object in the GW190814 event was most likely a low-mass black hole, confirming
our earlier conclusion~\cite{Sedrakian2020}. Nevertheless, a neutron star
interpretation cannot be excluded at this time, but would require a range
of extreme assumptions: (a) rapid (Keplerian) rotation, which may not be
reached due to various instabilities that may set in at lower rotation
frequencies; (b) strongly attractive $\Delta$-resonance potential in symmetric
nuclear matter; (c) large, positive value of the isoscalar skewness
$Q_{\mathrm{sat}}$ parameter.

\section*{Acknowledgements}
This work was, in part, supported by European COST Actions ``PHAROS"
(CA16214). The research of J.J.L. at Goethe University was supported by
the Alexander von Humboldt foundation. A.S. is supported by the
Deutsche Forschungsgemeinschaft (Grant No. SE 1836/5-1) and F.W. is
supported through the U.S.\ National Science Foundation under Grants
PHY-1714068 and PHY-2012152.

\bibliographystyle{elsarticle-num}
\bibliography{GW19_refs}
  
\end{document}